\documentclass{PoS}

\title{Dispersive analysis of $\omega/\phi\to3\pi$ decays and the $\omega/\phi\to\pi^0\gamma^*$ transition form factors}

\ShortTitle{Dispersive analysis of $\omega/\phi\to3\pi$ and $\omega/\phi\to\pi^0\gamma^*$}

\author{\speaker{Sebastian P. Schneider},{$^{a}$} Bastian Kubis, and Franz Niecknig\\
        Helmholtz-Institut f\"ur Strahlen- und Kernphysik (Theorie), Bethe Center for Theoretical Physics, Universit\"at Bonn, D-53115 Bonn, Germany\\
        \llap{$^a$}E-mail: \email{schneider@hiskp.uni-bonn.de}}

\abstract{We present a dispersive analysis of the three-pion decays of the lightest isoscalar vector mesons. The framework allows for a consistent implementation of final-state interactions among 
	  all three pions, with the results being solely dependent on the pion-pion P-wave scattering phase shift. We compare our results with the very precise KLOE measurement of $\phi\to3\pi$. 
	  The $\omega/\phi\to3\pi$ partial-wave amplitude serves as input to a dispersive analysis of the $\omega/\phi\to\pi^0\gamma^*$ transition form factor. We compare to data for $\omega\to\pi^0\mu^+\mu^-$
	  by the NA60 collaboration and motivate an experimental measurement of the OZI forbidden $\phi\to\pi^0\ell^+\ell^-$ decays in order to understand the strong deviations in the $\omega$ channel.}

\FullConference{The 7th International Workshop on Chiral Dynamics,\\
		August 6 -10, 2012\\
		Jefferson Lab, Newport News, Virginia, USA}

\begin{document}

\section{Introduction}

Due to its relative simplicity, the decays of the lightest isoscalar vector mesons into three pions are an ideal testing ground for a dispersion-theoretical approach. Indeed, due to Bose symmetry only
odd partial waves can contribute to the decay, and neglecting the discontinuities of F- and higher partial waves the system is completely determined by the $\pi\pi$ P-wave phase shift, which is
known very precisely~\cite{Bern,Madrid}. On the experimental side a large existing data base has already been established for $\phi\to3\pi$ by the KLOE and CMD-2 collaborations~\cite{KLOE,CMD2}, while
plans for Dalitz plot analyses of the $\omega\to3\pi$ channel at WASA and KLOE are well progressed. Moreover, the $\phi\to3\pi$ decay allows us to study the effects of crossed-channel rescattering 
effects on resonances, namely the $\rho$, in the decay region.

The $\omega/\phi\to\pi^0\gamma^*$ transition form factors have garnered considerable attention recently due to their possible impact on hadronic light-by-light scattering, which is believed to soon constitute 
the dominant uncertainty in $(g-2)_\mu$. They are expected to help constrain the pseudoscalar pole terms, whose strength is determined by the doubly-virtual form factor 
$F_{\pi^0\gamma^*\gamma^*}(M_{\pi^0}^2,q_1^2,q_2^2)$, for fixed isoscalar photon virtualities, $q_1^2=M_\omega^2$. We will use the $\omega/\phi\to3\pi$ partial-wave amplitude and the precisely known 
pion vector form factor as input to our calculations. Data for $\omega\to\pi^0\ell^+\ell^-$ has been collected by several groups. We will concentrate on measurements presented in Refs.~\cite{LeptonG,NA60,NA60new}.

\section{Dispersion relations for $\omega/\phi\to3\pi$}

\begin{figure}[t]
  \includegraphics[width= 0.47\linewidth]{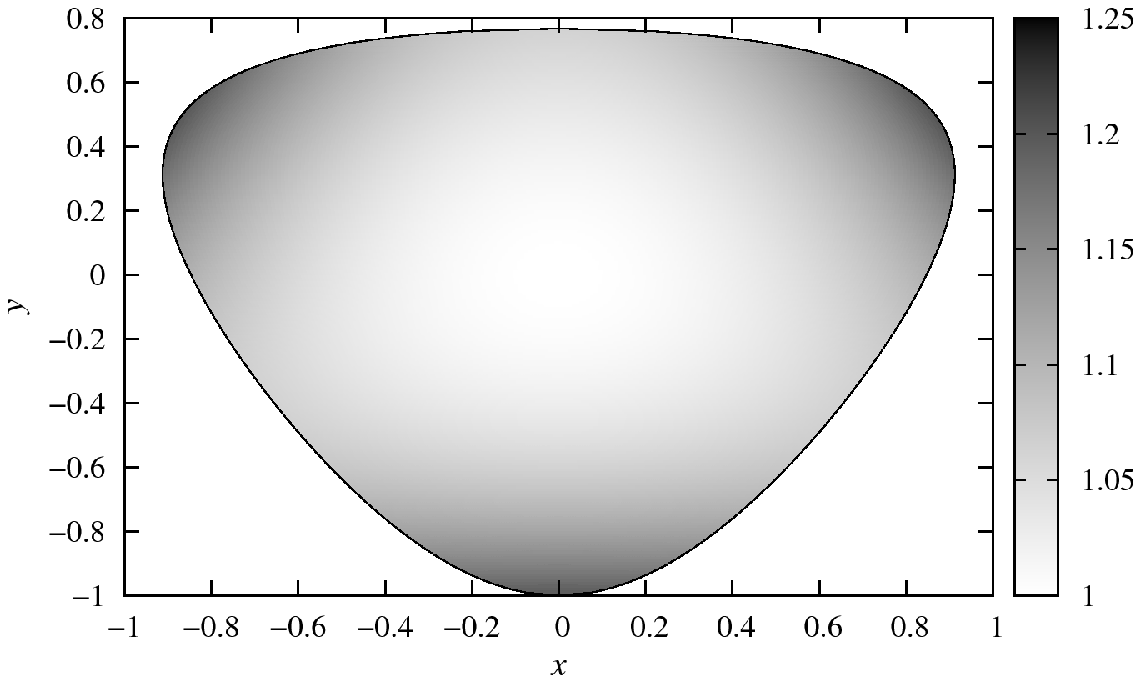}
  \includegraphics[width= 0.47\linewidth]{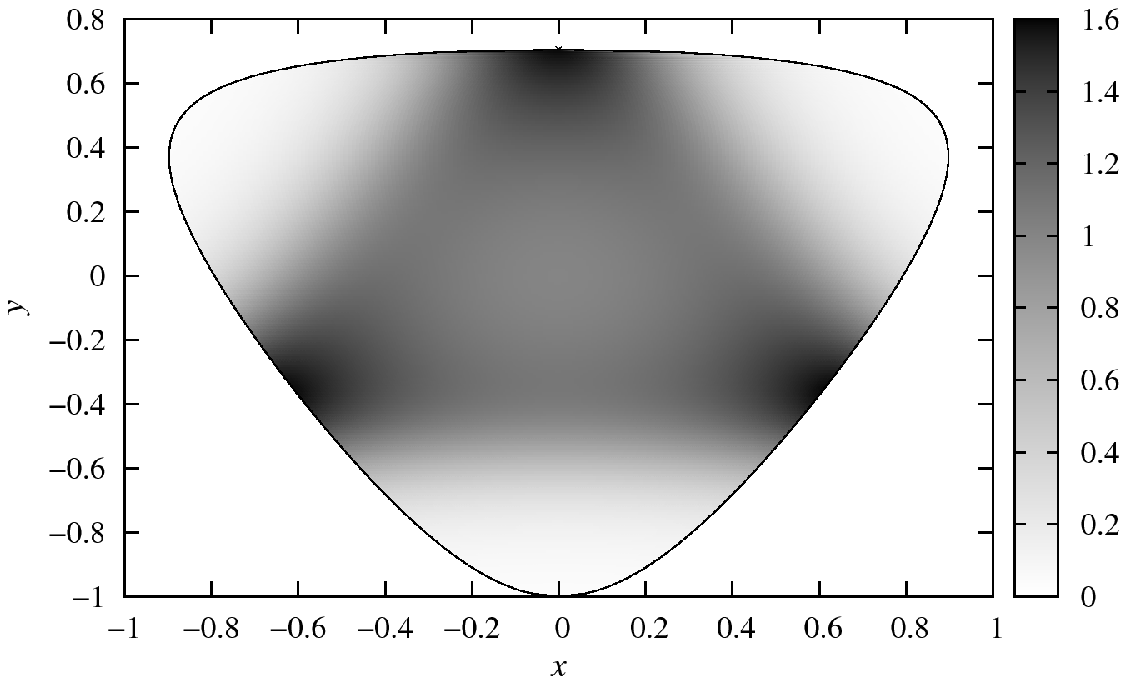}
  \caption{Dalitz plots for $\omega\to3\pi$ (left panel) and $\phi\to3\pi$ (right panel), divided by the P-wave phase space and normalized to one in the center.}
\label{fig:Dalitzomegaphi}
\end{figure}

A dispersion theory approach to hadronic three-body decays is not a new subject. In fact, it has been used in the past to analyze the $\omega\to3\pi$ decay~\cite{AitchisonGolding}. We will resort to the method
devised in Ref.~\cite{AnisovichLeutwyler} for $\eta\to3\pi$. The integral equations derived for numerical calculations are based on fundamental physics principles: unitarity, analyticity and crossing symmetry.
The only approximation that enters our calculations is the assumption of elastic $\pi\pi$ scattering and that discontinuities of F waves are small enough to be neglected (this issue is more closely
investigated in Ref.~\cite{V3pi}). In that approximation the scattering amplitude can be decomposed into functions of one variable with only a right-hand cut,
\begin{equation}
 \mathcal{F}(s,t,u)=\mathcal{F}(s)+\mathcal{F}(t)+\mathcal{F}(u)~,
\end{equation}
where $s,t,u$ are the Mandelstam variables of the decay and $s+t+u=M_{\omega/\phi}^2+3M_\pi^2=3s_0$. The functions $\mathcal{F}(s)$ fulfill the unitarity relation
\begin{equation}\label{eq:unrel}
 {\rm disc}\mathcal{F}(s)=2i\theta(s-4M_\pi^2)\{\mathcal{F}(s)+\hat\mathcal{F}(s)\}\sin\delta(s)e^{-i\delta(s)}~,
\end{equation}
where $\delta(s)$ is the $\pi\pi$ P-wave phase shift and the inhomogeneity $\hat\mathcal{F}(s)$ is given as
\begin{equation}\label{eq:inhom}
 \hat\mathcal{F}(s)=3\langle(1-z^2)\mathcal{F}\rangle(s)~,\qquad \langle z^nf\rangle(s)=\frac{1}{2}\int_{-1}^1dz\,z^nf\Bigl(\frac{1}{2}\bigl(3s_0-s+z\kappa(s)\bigr)\Bigr)~,
\end{equation}
with $\kappa(s)=\sigma_\pi(s)\lambda^{1/2}(M_{\omega/\phi}^2,M_\pi^2,s)$, $\sigma_\pi(s)=\sqrt{1-4M_\pi^2/s}$, and the K\"all\'en function $\lambda(x,y,z)=x^2+y^2+z^2-2(xy-xz-yz)$. Some care has to
be taken to perform the angular integration to respect the analytic structure of the decay, we point to Ref.~\cite{V3pi} for further details. A solution to Eq.~(\ref{eq:unrel}) is given by an 
integral equation,
\begin{equation}\label{eq:inteq}
 \mathcal{F}(s)=\Omega(s)\biggl\{a+\frac{s}{\pi}\int_{4M_\pi^2}^\infty\frac{ds'}{s'}\frac{\sin\delta(s')\hat\mathcal{F}(s')}{|\Omega(s')|(s'-s)}\biggr\}~,
\end{equation}
where $a$ is the only subtraction constant of the system and 
\begin{equation}
 \Omega(s)=\exp\biggl\{\frac{s}{\pi}\int_{4M_\pi^2}^\infty ds'\frac{\delta(s')}{s'(s'-s)}\biggr\}
\end{equation}
is the Omn\`es function~\cite{Omnes}. The integral equation~(\ref{eq:inteq}) is solved together with Eq.~(\ref{eq:inhom}) by an iterative numerical procedure. In Fig.~\ref{fig:Dalitzomegaphi} we display
the Dalitz plot, divided by phase space and normalized to one in the center, in terms of the variables
\begin{equation}
 x=\frac{\sqrt{3}(t-u)}{2\,M_{\omega/\phi}(M_{\omega/\phi}-3M_\pi)}~, \qquad y=\frac{3(s_0-s)}{2\,M_{\omega/\phi}(M_{\omega/\phi}-3M_\pi)}~.
\end{equation}
Performing a trivial redefinition, we can show that $a$ serves as the overall normalization of the decay amplitude~\cite{V3pi}, so that the \emph{normalized} Dalitz plot depends \emph{only} on the input for the $\pi\pi$ P-wave phase shift, for which we use the results from Refs.~\cite{Bern,Madrid}. The $\omega\to3\pi$ Dalitz plot shows a 
smooth distribution, which rises from the center to the outer borders. The available phase space is obviously not sufficient to contain the $\rho$ resonance, nevertheless it fixes the sign of the
leading slope parameter of the Dalitz plot parameterization unambiguously. The $\phi\to3\pi$ Dalitz plot shows significantly more structure, since the physical region encompasses 
the $\rho$ resonance. From its center, the Dalitz plot distribution rises towards these bands and then steeply falls off, showing almost complete depletion towards the outer corners.

\begin{figure}[t]
\centering
\includegraphics[width = 0.49\linewidth]{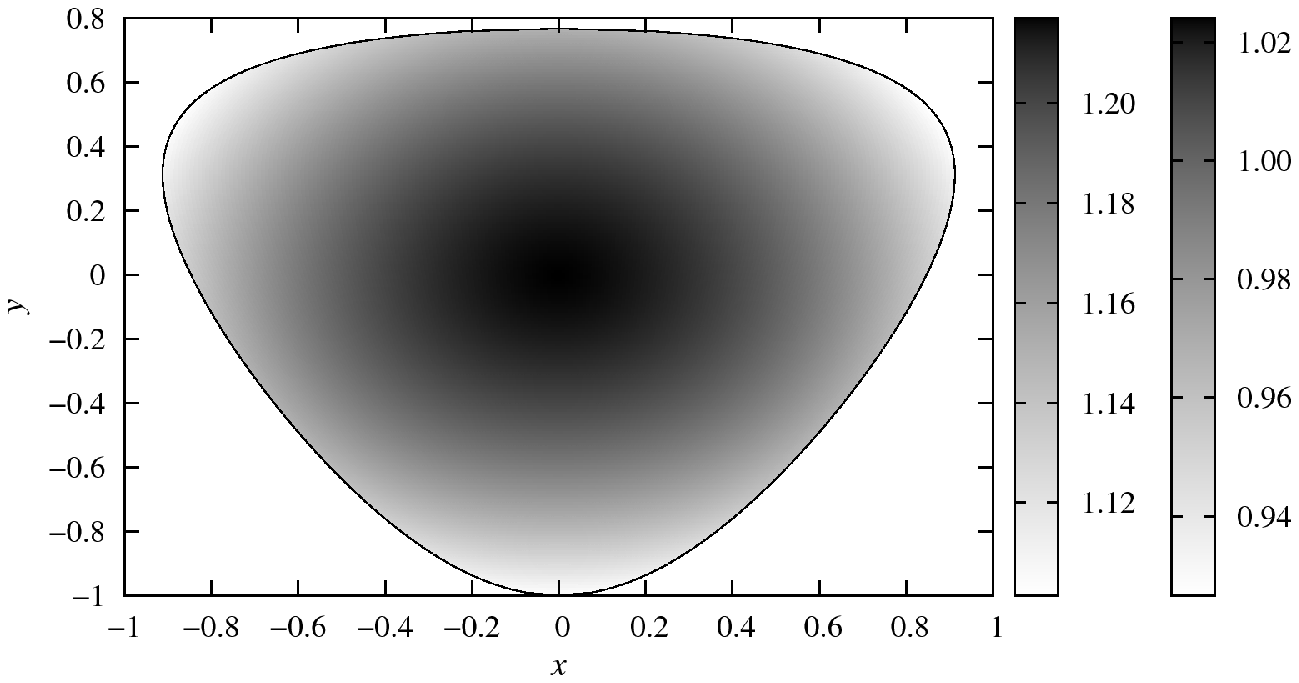}
\includegraphics[width = 0.49\linewidth]{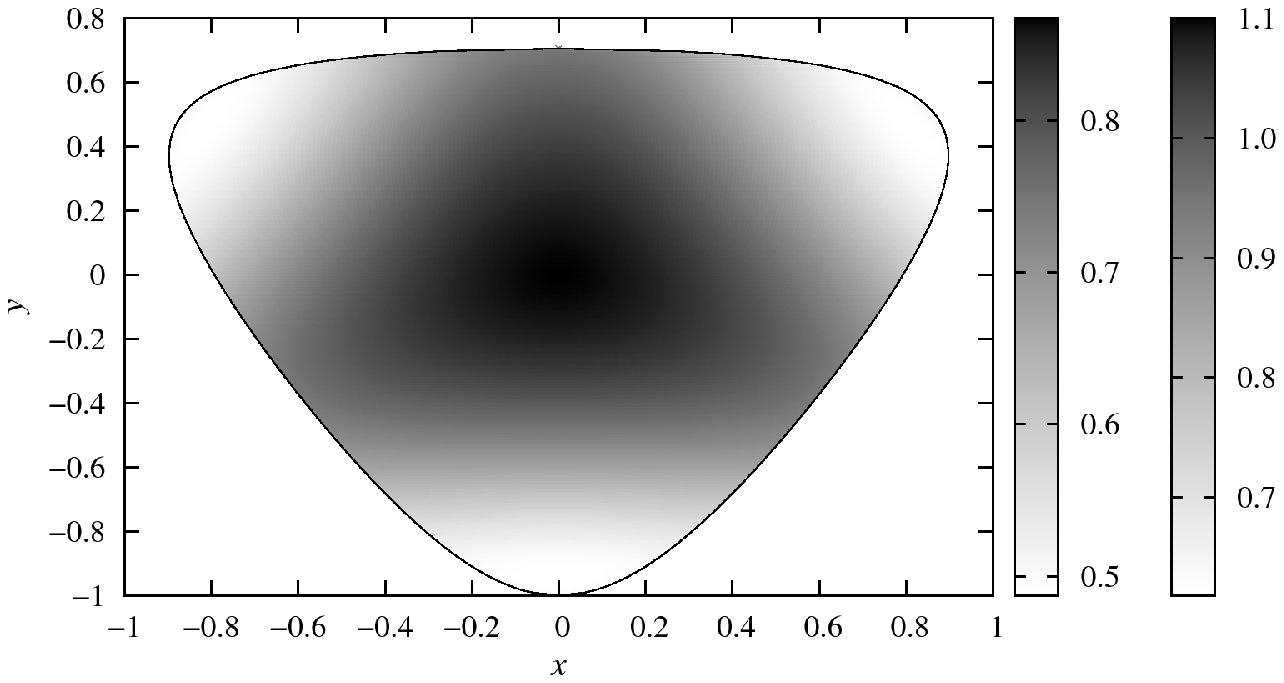} 
\caption{$|\mathcal{F}_{\rm full}|^2/|\mathcal{F}_{\hat\mathcal{F}=0}|^2$ for $\omega\to3\pi$ (left panel) and $\phi\to3\pi$ (right panel), see text for details.}
\label{fig:crossomegaphi}
\end{figure}

It remains to analyze the influence of crossed-channel rescattering. This is done in Fig.~\ref{fig:crossomegaphi}. We devise two separate methods to fix the subtraction constant $a$. For the left-hand
scale of either diagram $a$ is matched to the decay rate, $\Gamma_{\omega\to3\pi}=7.56$ MeV and $\Gamma_{\phi\to3\pi}=0.65$ MeV respectively, before the iteration. We observe that in that case the 
partial width of the final result is increased by about 20\% for $\omega\to3\pi$ and decreased by the same amount for $\phi\to3\pi$. For the right-hand scales we match $a$ to the decay rate before and 
after the iteration and observe that a significant part of the modifications is absorbed in an overall normalization. We also observe that the $\rho$ resonance bands are left relatively unaffected.

\begin{figure}[b]
\centering
\includegraphics*[width = 0.45\linewidth]{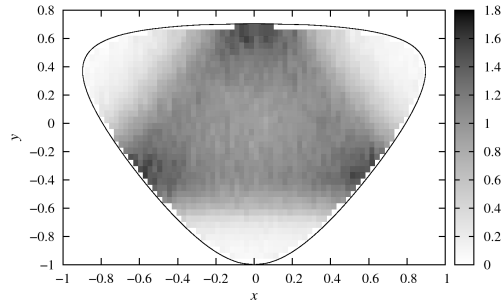}
\caption{Dalitz plot measured by the KLOE collaboration~\cite{KLOE}. Shown is the efficiency-corrected number of counts in the respective bin, divided by phase-space and normalized to 1 in the Dalitz plot center.}
\label{fig:KLOE}
\end{figure}

Comparing to the experimental $\phi\to3\pi$ Dalitz plot, we fit the subtraction constant as an overall normalization to the KLOE data shown in Fig.~\ref{fig:KLOE}. The $\chi^2/{\rm ndof}$ of the fit 
without rescattering effects, $\hat\mathcal{F}=0$, ranges between $1.71\ldots2.06$, while the uncertainty range stems from varying between different phase shift parameterizations and uncertainties 
related to their high-energy behavior~\cite{V3pi}. Rescattering effects improve the $\chi^2/{\rm ndof}$ to $1.17\ldots 1.50$. While this points to an overall 
improvement of the fit quality with rescattering effects, the goodness of the fit is still rather low due to the high precision of the KLOE data. 

We can oversubtract the dispersion integral in Eq.~(\ref{eq:inteq}) at the expense of predictability and the correct high-energy behavior of the amplitude $\mathcal{F}(s)$. This leads to
\begin{equation}
 \mathcal{F}(s)=\Omega(s)\biggl\{a+bs+\frac{s^2}{\pi}\int_{4M_\pi^2}^\infty\frac{ds'}{s'^2}\frac{\sin\delta(s')\hat\mathcal{F}(s')}{|\Omega(s')|(s'-s)}\biggr\}~.
\end{equation}
If we then include the additional subtraction constant $b$ in our fit, we find a perfect representation of the data with $\chi^2/{\rm ndof}=1.02\ldots1.03$. Note that this representation respects unitarity, 
analyticity and crossing symmetry. It can be used as input for the $\omega/\phi\to\pi^0\gamma^*$ partial-wave amplitude in an analysis of the $\omega/\phi\to\pi^0\gamma^*$ transition 
form factor. Further details on the above discussions may be found in Ref.~\cite{V3pi}.

\section{Dispersion relations for the $\omega/\phi\to\pi^0\gamma^*$ transition form factor}

Assuming that the $\omega/\phi\to\pi^0\gamma^*$ transition form factor $f_{V\pi^0}(s)$ ($V=\omega/\phi$) is dominated by $\pi\pi$ intermediate states\footnote{Due to the fact that $\gamma^*$ is required to be an isovector photon, this is a good approximation.}
one can derive the unitarity relation (see also~\cite{Koepp})
\begin{equation}
 {\rm disc} f_{V\pi^0}(s)=\frac{i s}{48\pi}\sigma_\pi^3(s)f_1(s)F_\pi^{V*}(s)~,
\end{equation}
where $f_1(s)=\mathcal{F}(s)+\hat\mathcal{F}(s)$ is the previously determined $V\to3\pi$ partial-wave amplitude and $F_\pi^{V*}(s)$ is the well-known pion vector form factor.
This relation leads to a once-subtracted dispersion relation
\begin{equation}\label{eq:TFF}
 f_{V\pi^0}(s)=f_{V\pi^0}(0)+\frac{s}{96\pi^2}\int_{4M_\pi^2}^\infty ds'\frac{\sigma_\pi^3(s')f_1(s')F_\pi^{V*}(s')}{s'-s}~,
\end{equation}
where the subtraction constant is fixed by the real-photon partial width $\Gamma_{V\to\pi^0\gamma}$. In principle the asymptotic behavior of the partial-wave amplitude and the pion vector form factor
even allows for an unsubtracted dispersion relation, but the above is numerically much more stable. We have calculated $\Gamma_{V\to\pi^0\gamma}$ by a sum rule for $f_{V\pi^0}(s)$ and find that it is 
saturated to about 90--95\% by two-pion intermediate states, thus justifying the approximation of neglecting inelastic intermediate states. 

\begin{figure*}[t]
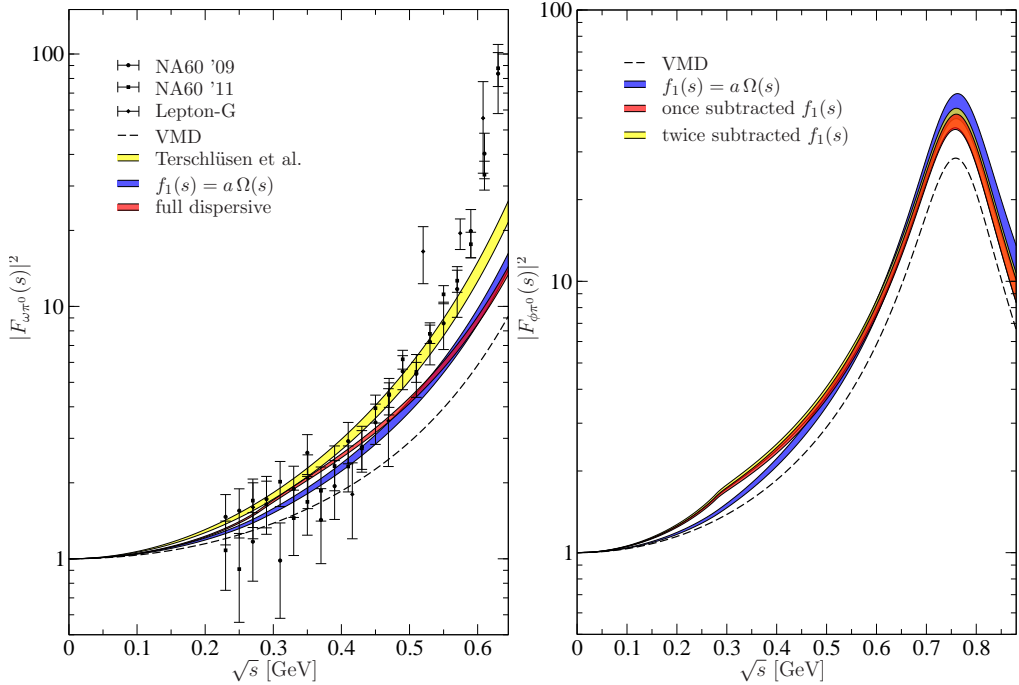

\centering
\includegraphics[clip=true,width=0.44\linewidth]{TFF.eps}
\includegraphics[clip=true,width=0.44\linewidth]{TFFPhi.eps}
\caption{Left: numerical results for $\omega\to\pi^0\gamma^*$. We show pure VMD (dashed line), the results of a chiral Lagrangian treatment with explicit vector mesons~\cite{Terschlusen1} (yellow band),
the dispersive solution for $f_1(s)=a\,\Omega(s)$ (blue band), and the full dispersive solution with one subtraction in the $V\to3\pi$ partial-wave amplitude (red band). Right: Numerical results for 
$\phi\to\pi^0\gamma^*$. Again, we show pure VMD (dashed line), the dispersive solution for $f_1(s)=a\,\Omega(s)$ (blue band), and the full dispersive solution with one subtraction (red band) and two 
subtractions (yellow band).}
 \label{fig:TFF}
\end{figure*}

In Fig.~\ref{fig:TFF} we display the numerical results for the normalized transition form factor $F_{V\pi^0}(s)=f_{V\pi^0}(s)/f_{V\pi^0}(0)$. Although we significantly improve the
vector-meson dominance (VMD) result, we cannot accommodate the steep rise in experimental data. In fits of the transition form factor this structure is often represented by a pole term at around 
$600$--$700$ MeV, the physical nature of which is unknown. We also find that three-particle effects in the partial-wave amplitude do not perturb the spectrum in a way which is observable at the current precision level of data. 
The only loose end at this stage of the discussion is the fact that our $\omega\to3\pi$ partial-wave amplitude is not backed up by experimental data.

This problem is remedied when considering the $\phi\to\pi^0\gamma^*$ transition form factor. The twice-subtracted partial-wave amplitude is an extremely precise representation of data, and thus all input
in this channel is well constrained. Our numerical results again show enhancement over VMD, while crossed-channel rescattering effects are not particularly strong. We note that the physical region of
the decay encompasses the $\rho$ resonance, but there is no indication for any particular behavior of the transition form factor as suggested by $\omega\to\pi^0\mu^+\mu^-$ data. Based on these 
observations we strongly suggest an experimental investigation of $\phi\to\pi^0\ell^+\ell^-$: it is reasonable to assume that the steep rise at around $600$--$700$ MeV also occurs in this channel, with
the added advantage that in this case it would be part of the observable region and may thus significantly help to improve our understanding of transition form factors.

A more detailed discussion of the $V\to\pi^0\gamma^*$ transition form factor is given in Ref.~\cite{TFF}.

\end{document}